\documentstyle[12pt]{article}

\def\rdots{\mathinner{\mkern1mu\raise1pt\vbox{\kern1pt\hbox{.}}\mkern2mu
   \raise4pt\hbox{.}\mkern2mu\raise7pt\hbox{.}\mkern1mu}}
\newcommand{\be}{\begin{equation}}
\newcommand{\ee}{\end{equation}}

\newcommand{\Z}{{\rm Z\kern-.35em Z}}
\newcommand{\bP}{{\rm I\kern-.15em P}}
\newcommand{\Q}{\kern.3em\rule{.07em}{.65em}\kern-.3em{\rm Q}}
\newcommand{\R}{{\rm I\kern-.15em R}}
\newcommand{\h}{{\rm I\kern-.15em H}}
\newcommand{\C}{\kern.3em\rule{.07em}{.65em}\kern-.3em{\rm C}}
\newcommand{\T}{{\rm T\kern-.35em T}}
\newcommand{\D}{{\kern-.5em }}

\begin{document}

\openup 1.5\jot
\centerline{A Criterion for Ferromagnetism in the Isotropic Quantum
Heisenberg Model}

\vspace{1in}
\centerline{Paul Federbush}
\centerline{Department of Mathematics}
\centerline{University of Michigan}
\centerline{Ann Arbor, MI 48109-1109}
\centerline{(pfed@umich.edu)}

\vspace{2in}

\centerline{\underline{Abstract}}

A criterion for ferromagnetism is presented suggesting a line of proof to
rigorously establish the phase transition.  Spectral information will be
required in certain invariant subspaces of the Hamiltonian, hopefully the
relatively crude estimates needed will be not too difficult to establish.

\vfill\eject

We hope the criterion presented in this paper leads in short order to a
rigorous proof of the ferromagnetic phase transition for the isotropic
quantum Heisenberg model.

The model is constructed on a rectangular lattice, $V$.  The Hamiltonian
is taken as
\be     H = - \sum_{i \sim j} \frac 1 2 \Big( \vec \sigma_i \cdot \vec
\sigma_j - 1 \Big) = - \sum_{i \sim j}\Big( I_{ij}-1\Big) \ee

where $I_{ij}$ interchanges the spins at nearest neighbor sites $i$ and
$j$.  For the Hilbert space we introduce an o.n. basis with elements

\be
\vec i_{{\cal S}} =
\begin{array}[t]{c}
{\displaystyle\otimes} \\
{\scriptstyle {i\in {\cal S}}}
\end{array}
\left( \begin{array}{c} 1 \\ 0 \end{array} \right)_i
\begin{array}[t]{c}
{\displaystyle\otimes} \\
{\scriptstyle {j \not\in {\cal S}} }
\end{array}
\left( \begin{array}{c} 0 \\ 1 \end{array} \right)_j
\ee
with each $\cal S$ a subset of the vertices.  The $r$ spin wave sector is
spanned by the $\vec i_{{\cal S}}$ with $\#({\cal S})=r$, and is an
invariant subspace of $H$.

In [1] operators $T^{r,s}$ were introduced mapping the $r$ spin wave
sector of the Hilbert space to the $s$ spin wave sector.  If

\be     \vec g = \sum_{{\cal S}} g({\cal S}) \vec i_{\cal S}    \ee
with the cardinality of all sets in this sum equal $r$.  Then
\be \vec h = T^{r,s} \vec g = \sum_{\cal S} h({\cal S}) \vec i_{\cal S} 
\ee
where in this sum all sets are of cardinality $s$, and
\be     h({\cal S}) = \sum_{{\cal S}' \supset {\cal S}} g({\cal S}') \ \
{\rm if} \;\#({\cal S}) = s.   \ee
It was shown in [1] that these operators intertwine with $H$,
\be     HT^{r,s} = T^{r,s} \ H, \ee
(really a trivial property due to invariance under global rotation).

We will make some arbitrary choices in our present exposition, for
simplicity.   We let lattice size become infinite through a sequence of
lattices all with, $v =  |V|$, the number of their vertices, divisible by
10.  We decompose the trace into contributions from the different spin
wave sectors.

\be
Tr \left( e^{-\beta H}\right)_V \equiv Tr(V,\beta) = \sum^v_{i=0}
Tr(V,\beta,i)
\ee
where in Tr$(V, \beta, i)$ only basis elements, $\vec i_{{\cal S}}$, with
$\#({\cal S}) = i$ are kept.  We use the following criterion for
permanent magnetism, derived from Section 2 of [2].

\noindent

\underline{Magnetism Criterion}:  The system is ferromagnetic for $\beta
\ge \beta_0$ if for some $M_0$, and $v \ge M_0$ one has

\[      Tr(V, \beta,i) \le 2 \ Tr(V, \beta, i - \frac{v}{10} )  \]
\be \ \ \ \ \ \ \ \ \ {\rm for} \ \ \ i = \frac 4{10} \ v + 1, ....,
\frac v 2  \ee
when $\beta \ge \beta_0$ and $v \ge M_0$.

Our hope for an attack on the phase transition problem is to use the
operators $T^{i,i-\frac{v}{10}}$  to study the relation between
Tr$(V,\beta,i)$ and Tr$(V, \beta, i - \frac{v}{10})$.  If we let $K_i$ be
the kernel of $T^{i,i-\frac{v}{10}}$ and $R_i$ be the orthogonal
complement of $K_i$ in the $i$ spin wave sector, and gets

\be     Tr(V,\beta,i) = Tr^{(1)} (V,\beta,i) + Tr^{(2)}(V, \beta,i)    
\ee
where $Tr^{(1)}$ is the portion of the trace from $K_i$ and $Tr^{(2)}$
the portion from $R_i$.  One has
\be     Tr^{(2)}(V,\beta,i) = Tr(V, \beta, i - v/10)    \ee
and one is left with trying to prove
\be     Tr^{(1)}(V,\beta,i) \le Tr(V, \beta, i - v/10)  \ee
(for $\beta \ge \beta_0, \ v \ge M_0$).
For the range of $i$ needed, relation (10) is easily shown.  We have
ideas on how to develop a proof of  (11).

The reduction of this paper, from proving the phase transition to
studying (11), is truly trivial.  But we believe it is new.  If this has
been known to someone, likely then for a long time, perhaps they've
missed the boat in not pushing it through to a proof.

\vfill\eject

\centerline{\underline{References}}

\bigskip

\begin{itemize}

\item[[1]] P. Federbush, ``A Polymer Expansion for the Quantum Heisenberg
Wave Function", to be published in {\it J. Math. Phys.}, math-ph/0302067.

\item[[2]]  P. Federbush, ``For the Quantum Heisenberg Ferromagnet, Tao
to the Proof of a Phase Transition",   math-ph/0202044.

\end{itemize}

\end{document}